# Spin Seebeck in the weak exchange coupled van der Waals antiferromagnet

Xue He[1], Shilei Ding[2*], Hans Gløckner Giil[3], Jicheng Wang[1], Zhongchong Lin[4], Zhongyu Liang[4], Jinbo Yang[4*], Mathias Kläui[3,5], Arne Brataas[3], Yanglong Hou[6,7*], Rui Wu[1*]

1. Spin-X Institute, School of Physics and Optoelectronics, State Key Laboratory of Luminescent Materials and Devices, and Guangdong-Hong Kong-Macao Joint Laboratory of Optoelectronic and Magnetic Functional Materials, South China University of Technology, Guangzhou 511442, China

2. Department of Materials, ETH Zürich, 8093 Zürich, Switzerland

3. Center for Quantum Spintronics, Norwegian University of Science and Technology, Trondheim 7491, Norway

4. State Key Laboratory for Mesoscopic Physics, School of Physics, Peking University, Beijing 100871, P.R. China

5. Institute of Physics, Johannes Gutenberg-University Mainz, Staudingerweg 7, Mainz 55128, Germany

6. School of Materials, Shenzhen Campus of Sun Yat-Sen University, Shenzhen 518107, China

7. School of Materials Science and Engineering, Beijing Key Laboratory for Magnetoelectric Materials and Devices, Peking University, Beijing 100871, China

*Corresponding author: shilei.ding@mat.ethz.ch, jbyang@pku.edu.cn, hou@sysu.edu.cn, and ruiwu001@scut.edu.cn.

**Spin Seebeck effect (SSE) refers to the creation of spin currents due to a temperature gradient in the magnetic materials or across magnet-normal metal interfaces, which can be electrically detected through the inverse spin Hall effect (ISHE) when in contact with heavy metals. It offers fundamental insights into the magnetic properties of materials, including the magnetic phase transition, static magnetic order, and magnon excitations. However, the SSE in van der Waals antiferromagnet is still elusive, especially across the spin-flip transition. Here, we demonstrate the SSE in the weak exchange coupled van der Waals antiferromagnet**

**CrPS$_4$. The SSE increases as the magnetic field increases before the spin-flip transition due to the enhancement of the thermal spin current as a function of the applied field. A peak of SSE is observed at the spin-flip field, which is related to the magnon mode edges across the spin-flip field. Our results extend SSE research to van der Waals antiferromagnets and demonstrate an enhancement of SSE at the spin-flip transition.**

Thermoelectricity combines heat transfer and electric voltage in solid materials, presenting a promising option for green energy production by harnessing waste heat with a simple device design[1]. In particular, thermal spintronics effect utilize nonequilibrium magnon transport phenomena in the presence of a heat gradient, enabling magnetic insulators to serve as effective thermoelectric devices[2]. The spin Seebeck effect (SSE) has therefore drawn significant interest, where a temperature gradient ($\nabla T$) in magnetic materials leads to the generation of spin currents ($J_s$). SSE can be subsequently detected via the inverse spin Hall effect (ISHE) in a heavy metal contact with strong spin-orbit coupling[3-24]. The phenomenon was initially discovered in 2008[5], and various configurations have been suggested to explore the SSE, such as transverse SSE[6], longitudinal SSE[7], and nonlocal SSE[8]. Additionally, it has been examined in a range of magnetically ordered systems, including ferromagnets[5,9], ferrimagnets[6,10], antiferromagnets[11-15], paramagnets[16], chiral helimagnets[17], and quantum magnets[18], where the magnon excitations play critical roles regardless of long-range or short-range magnetic interactions.

In ferromagnet/heavy metal bilayers, the SSE observed below the Curie temperature is associated with the spin current generated by thermally excited magnons that exhibit only right-handed chirality[19]. The SSE mechanism in antiferromagnetic heterostructures is more complex due to two magnetic sublattices, which result in different magnon modes[20-23]. In a uniaxial antiferromagnet, there are two magnon branches with opposite chirality carrying opposite angular momentum. These modes are degenerate at zero magnetic field, meaning there is no net magnon current until a field is applied to lift this degeneracy. A change in the sign of the SSE was observed during the spin-flop transition[14,15], which is attributed to the change in the chirality of the thermally excited magnon mode, which dominates. Additionally, the interfacial Néel coupling and spin conductance can influence the sign and magnitude of the SSE[21,23]. Nonetheless, the spin Seebeck effect in van der Waals antiferromagnets requires further investigation[25,26], particularly in the van der Waals system with interlayer antiferromagnetic coupling, which typically suggests weak exchange coupling and a low spin-flip field.

CrPS$_4$ is an antiferromagnetic van der Waals material constituted of chains of chromium octahedra interconnected through phosphorus[27-33] as shown in Fig. 1a. Due to the chemical composition and multi-bonded crystal structures, CrPS$_4$ is a comparably air-stable material that makes the device fabrication easier compared with other van der Waals materials[34]. It shows a sizeable Néel temperature ($T_\mathrm{N} = 36$ K) and A-type antiferromagnetic ordering[27]. Unlike the conventional bulk antiferromagnetic materials, CrPS$_4$ with a layered structure exhibits extremely weak interlayer interactions between sublattice spins, where spins within each monolayer are aligned ferromagnetically out of the plane, subsequently leading to the weak spin-flop field (0.8 T at 5 K) and spin-flip field (7 T at 5 K) as shown in Fig. 1b. This characteristic also significantly lowers the frequency of antiferromagnetic magnons to the GHz range[35]. As a result, it provides easier access to antiferromagnetic dynamics. Notably, it improves the efficiency of the thermal magnon population compared to traditional antiferromagnets with a large magnon gap, making CrPS$_4$ an excellent candidate for investigating the mechanism of SSE in antiferromagnets.

Here, we demonstrate the SSE in CrPS$_4$ in contact with a heavy metal. A vertical temperature gradient in CrPS$_4$ drives the magnon current in the longitudinal SSE configuration. The SSE increases as a function of the applied field before the spin-flip transition. The enhancement of the canted magnetization leads to pronounced magnon pumping. At the spin-flip field, a peak and saruration of SSE is observed that further disappears above the Néel temperature.

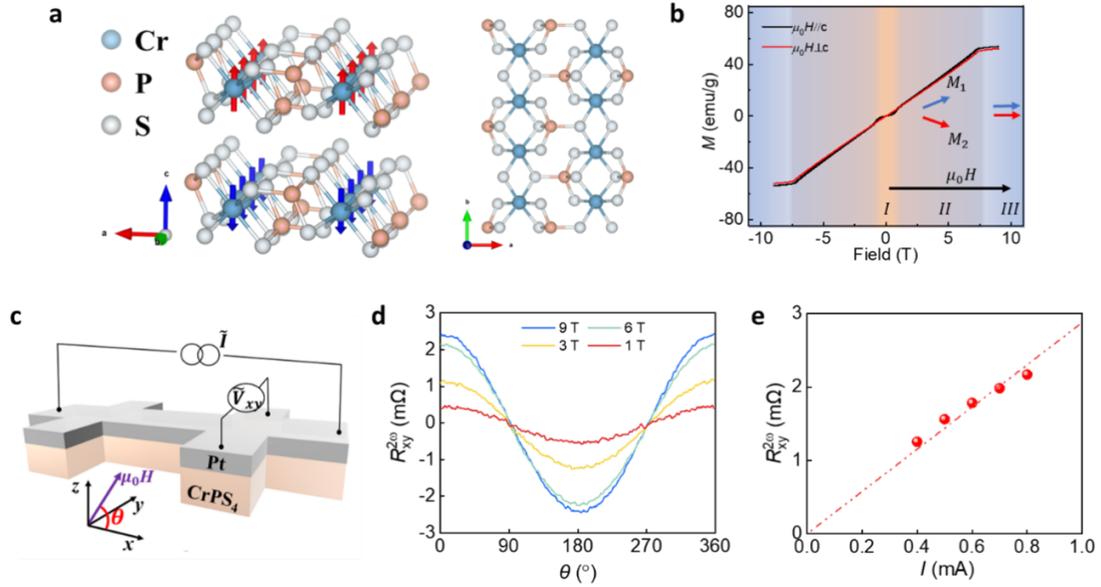

**Fig.1 Structure and magnetic properties of CrPS$_4$ and the Hall bar device for spin Seebeck effect (SSE) measurement**. **a** Crystal structure of CrPS$_4$. The red and blue arrows indicate the direction of the magnetic moment. **b** The magnetic measurements at 15 K are taken both along and perpendicular to the *c*-axis. The spin-flop and spin-flip transitions appear when the magnetic field is aligned with the *c*-axis. In contrast, only the spin-flip transition occurs when the field is applied perpendicularly to the *c*-axis, **c** Schematic of the Hall bar devices for the longitudinal spin Seebeck effect. The alternating current heats the sample, creating a vertical heat gradient and generating a spin current perpendicular to the sample plane. **d** Angular dependence (in the *xz* plane) of $R_{xy}^{2\omega}$ at different fields at a temperature of 15 K and an applied current of 1 mA (peak value). **e** Applied current dependence of $R_{xy}^{2\omega}$ (at 9 T) at 15 K. The dashed-dot line is the linear fit.

## Results

**Longitudinal SSE in CrPS$_4$/Pt(Ta)**. To obtain CrPS$_4$/Pt heterostructures for the SSE measurements, we deposited 5 nm Pt on top of exfoliated CrPS$_4$ flakes and subsequently fabricated Hall bar devices (see methods for details and schematic in Fig. 1c). The structure and phase of the CrPS$_4$ are characterized with X-ray Diffractometer and Raman spectroscopy (details see Supplemental Material S1). The microscopic picture of the CrPS$_4$/Pt Hall bar device can be found in Supplemental Material S2, where one could obtain the thickness of the CrPS$_4$ flake to be 75 nm. An alternating current ($\tilde{I}$) is applied to the Hall bar to generate vertical temperature gradient $\nabla T$, leading to the population of spin current $\boldsymbol{J}_s = -S\nabla T$. $S$ is the SSE coefficient. By further applying a magnetic field, it is possible to observe the SSE detected via the inverse spin Hall effect. The resultant electric field $\boldsymbol{E}_{\mathrm{ISHE}}$ is given by[3]

$$\boldsymbol{E}_{\mathrm{ISHE}} \propto \theta_{\mathrm{SH}} \boldsymbol{J}_s \times \boldsymbol{\sigma}, \tag{1}$$

Where $\theta_{\mathrm{SH}}$ is the spin Hall angle. $\boldsymbol{\sigma}$ is the spin polarization direction, which is parallel to the equilibrium magnetization $\boldsymbol{M}$. Since the temperature gradient results from the heating power of Pt, which is proportional to $\tilde{I}^2$, it is expected that the thermal signal can be detected through the second harmonic response $R_{\mathrm{xy}}^{2\omega}$.

In the magnetic material/Pt bilayer system, $R_{\mathrm{xy}}^{2\omega}$ typically involves different factors, including current-induced torque and thermal effects, which encompass the Nernst and spin Seebeck effects[36]. The electric field induced by the Nernst effect can be expressed as $\boldsymbol{E}_{\mathrm{NE}} \propto \nabla T \times \boldsymbol{M}$ [37], which share the same symmetry as SSE in the longitudinal configuration. When a strong magnetic field is applied, the current-induced torque is suppressed[36], leaving only thermal effects in the second harmonic response $R_{\mathrm{xy}}^{2\omega}$. Fig. 1d illustrates the angular dependence of $R_{\mathrm{xy}}^{2\omega}$ in the xz plane under different applied fields with the applied current of 1 mA (peak value) and the ambient temperature of 15 K. $R_{\mathrm{xy}}^{2\omega}$ reaches the maximum when the magnetic field is aligned with the x-axis and disappears when aligned with the z-axis (or c-axis), and the angular dependence data can be fitted well using the sine function following the Eq. (1). By applying an in-plane magnetic field, the Zeeman splitting lift the degeneracy of the two magnon eigenmodes, resulting in the spin current that induces the SSE signal (see discussion below and Fig. 3c). In the canted phase, the SEE increases with the strength of the applied magnetic field. This increase is generally attributed to the larger canted magnetization resulting from a strong magnetic field[23,38] or the increased SSE coefficient in response to the magnetic field[39]. This fundamentally differs from the SSE in ferromagnets, where an increased applied field would open the magnon gap, causing a decrease in SSE due to the reduction of the thermal magnon population[4]. Additionally, the magnitude of $R_{xy}^{2\omega}$ is proportional to the applied current as shown in Fig. 1e, demonstrating a thermoelectric nature similar to previous findings[40]. It is important to note that CrPS$_4$ has a semiconducting characteristic with an energy gap of $E_a = 0.166$ eV, and the resistivity ρ of CrPS$_4$ is reported to be ~ $2 \times 10^4$ Ω cm at 200 K[24]. By applying the Arrhenius equation[41] $\ln\rho = \ln\rho_0 - E_a/k_BT$, one could estimate that the resistivity of CrPS$_4$ below 50 K is higher than $1 \times 10^{12}$ Ω cm, allowing us to safely rule out the Nernst effect from conducting electrons in CrPS$_4$.

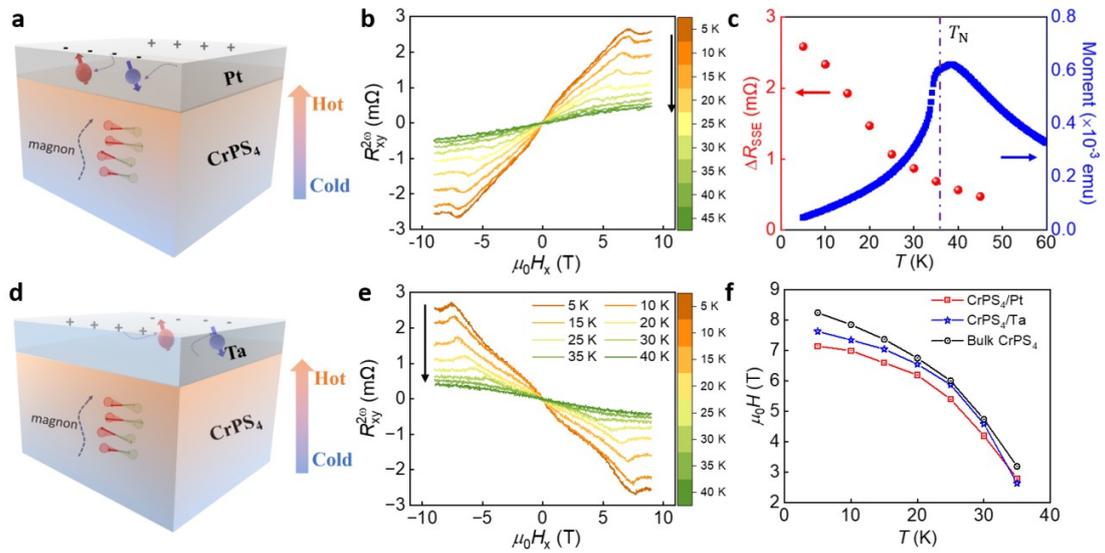

**Fig. 2 Temperature dependence of the SSE in CrPS$_4$/Pt and CrPS$_4$/Ta. a** and **d** The schematics of spin Seebeck effect in CrPS$_4$ in contact with Pt and Ta, the differing signs of the spin Hall angle result in a change in the sign of the SSE. **b** and **e** Field dependence ($\mu_0H_x$) of $R_{xy}^{2\omega}$ at various temperatures for both CrPS$_4$/Pt (5 nm) (applied current of 1 mA) and CrPS$_4$/Ta (11 nm) (applied current of 0.6 mA). **c** Temperature dependence of the SSE effective resistance in CrPS$_4$/Pt at 9 T, along with the magnetization as a function of temperature under a 50 mT applied field. The Néel temperature ($T_N$) is identified as 36 K, however, the SSE signal continues to be present even above $T_N$. **f** The field of the $R_{xy}^{2\omega}$ peak decreases with increasing temperature (blue star and red square), which is similar to the temperature dependence of the spin-flip transition field (black circle).

To better distinguish the SSE from other spurious effects, we utilize Pt and Ta in the two Hall bar devices (Fig. 2a and d). Due to the opposite spin Hall angles[42], the thermally generated spin current should yield SSE signals with opposite polarities in Pt and Ta samples. In contrast, other magnetic thermoelectric effects, such as the Nernst effect arising from the proximity effect[43], retain the same polarity in both Pt and Ta. As illustrated in Fig. 2b and e, the $R_{xy}^{2\omega}$ shows the opposite polarities in Pt and Ta samples, suggesting that the phenomenon originates from the SSE. As the temperature increases, the strength of the SSE decreases, and the SSE remains present even at temperatures exceeding the $T_N$ of $CrPS_4$. A more apparent trend is illustrated in Fig. 2c. Although the propagation of spin waves without magnetic interactions is not permitted in the paramagnetic phase, short-range magnetic interactions still facilitate short-wavelength magnetic excitations, resulting in the paramagnetic SSE[16]. In addition to the increase in $R_{xy}^{2\omega}$ with the applied field, peaks of $R_{xy}^{2\omega}$ are observed in both samples at varying temperatures. Similar effects are observed in the sample with a different Pt thickness (see Supplemental Material S3 for details). The magnetic field at which the $R_{xy}^{2\omega}$ peak occurs aligns with the spin-flip field of $CrPS_4$, as illustrated in Fig. 2f, suggesting a strong connection between the $R_{xy}^{2\omega}$ peak and the magnetic phase transition induced by the magnetic field. The longitudinal resistances for $CrPS_4$/Pt and $CrPS_4$/Ta are ~ 600 Ω and 1560 Ω respectively, with applied currents of 1 mA and 0.6 mA for the two samples. This results in a higher heating power in $CrPS_4$/Pt, causing a larger temperature difference between the sample and the chamber. There is expected to be a shift in the spin-flip field for the samples with and without heating at the same chamber temperatures, and this discrepancy will become more pronounced at lower temperatures (see Fig. 2f).

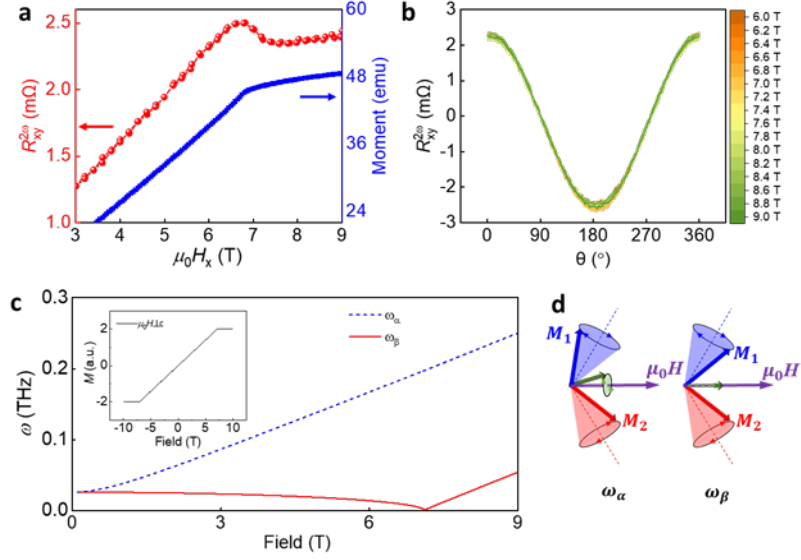

**Fig. 3 Origin of SSE peak at the spin-flip field. a** Comparison of the field dependence of $R_{xy}^{2\omega}$ in CrPS$_4$/Pt (obtained at 15 K) and magnetic moment CrPS$_4$ flake (measured at 20 K). **b** Angular dependence (in the $xz$ plane) of $R_{xy}^{2\omega}$ when the applied field approaches the spin-flip field at a temperature of 15 K and an applied current of 1 mA. **c** Magnon mode edges ($k = 0$) as a function of the applied field perpendicular to the $c$ axis. The inset shows the simulated magnetic moment as a function of the magnetic field. **d** The canted magnetization of $\omega_\alpha$ mode precesses around the applied field, while that of $\omega_\beta$ mode oscillates in the direction of the applied field.

**The origin of the SSE peak.** The peak of $R_{xy}^{2\omega}$ observed at the spin-flip field is intriguing as it is not associated with the static magnetic moment, which would not show an increased canted magnetization during the spin-flip transition, as illustrated in Fig. 3a. In Fig. 3b, the angular dependence (in the $xz$ plane) of $R_{xy}^{2\omega}$ is shown as the applied field approaches the spin-flip transition at a temperature of 15 K and an applied current of 1 mA in the CrPS$_4$/Pt. The curves can be well-fitted with the sine function according to Eq.(1), with a maximum observed at 6.8 T, indicating that the peak originates from the SSE. Additionally, the SSE continues to be present above $T_N$, while the peak of $R_{xy}^{2\omega}$ disappears beyond $T_N$ (see Fig. 2b,e). Although the paramagnetic phase could exhibit a SSE, the loss of long-range ordering above the $T_N$ causes the spin-flip transition to vanish.

This highlights the significant connection between the peak of SSE and the spin-flip transition.

The SSE consists of three components: 1. The temperature gradient excites the magnetization dynamics, leading to a non-equilibrium magnon current. 2. This magnon current is transformed into a conduction-electron spin current through the *s-d* interaction, which travels across the interface connected to the metal. 3. Finally, the spin current is converted into a charge current via the ISHE. Notably, detecting the spin current is not crucial for the SSE peak, as both the CrPS$_4$/Pt and CrPS$_4$/Ta samples exhibit peaks (see Fig. 2b,e). The only remaining likely mechanism for the SSE peak is related to the pumped spin current $J_s$ from the antiferromagnet into heavy metals which includes the effect of both thermal magnon excitation and interfacial spin mixing conductance.

Considering the canted magnetic phase, the magnetic field dependence of magnon frequency can be obtained by diagonalizing the spin Hamiltonian[44] with eigenfrequencies[45]. Before the spin-flip field, $\mu_0 H \leq 2\mu_0 H_E + \mu_0 H_A$,

$$\omega_\alpha = \gamma\mu_0 \sqrt{(2H_E \sin^2\varphi + H_A \cos^2\varphi)(2H_E + H_A)}, \qquad (2)$$

$$\omega_\beta = \gamma\mu_0 \sqrt{H_A(2H_E + H_A)\cos^2\varphi}, \qquad (3)$$

After the spin-flip field, $\mu_0 H > 2\mu_0 H_E + \mu_0 H_A$,

$$\omega_\alpha = \gamma\mu_0 \sqrt{H(H - H_A)}, \qquad (4)$$

$$\omega_\beta = \gamma\mu_0 \sqrt{(H - 2H_E)(H - 2H_E - H_A)}, \qquad (5)$$

where $\mu_0 H$, $\mu_0 H_E$, and $\mu_0 H_A$ represent the applied in-plane field, interlayer exchange field, and anisotropic field along the *c*-axis, respectively. The simplified model only considers the anisotropic field along the *c*-axis. $\omega_\alpha$ and $\omega_\beta$ are the two magnon modes. $\gamma$ is the gyromagnetic ratio and $\varphi$ is the canted angle along the *c*-axis applied in the plane field, $\varphi = \arcsin\frac{\mu_0 H}{2\mu_0 H_E + \mu_0 H_A}$.

The field dependence of the magnon mode frequency is plotted in Fig. 3c with parameters $\mu_0 H_E = 3.5$ T and $\mu_0 H_A = 0.12$ T [35]. The $\omega_\alpha$ mode has the potential to transport angular momentum due to the canted magnetization of the mode rotating around the applied magnetic field. This mode is similar to the quasi-ferromagnetic mode that emerges following a spin-flop transition when a magnetic field is applied along the *c*-axis[14]. Moreover, the SSE in CrPS4/Pt has the same sign as that in YIG/Pt (see Supplemental Material S4 for details), suggesting that right-handed magnons ($\omega_\alpha$ mode) are responsible for the SSE signal. In contrast, the $\omega_\beta$ mode oscillates in the direction of the applied field (see Fig. 3d).

We further calculate the spin current in the heavy metal following Ref. [23] using a minimal model where the CrPS4 sample is modeled as a one-dimensional antiferromagnetic chain with periodic boundary conditions. The model has an interfacial *s-d* coupling that couples the localized spins in the antiferromagnet with the itinerant electrons in the heavy metal. Using Fermi's Golden rule to calculate the transition probability for the spins to be pumped from the antiferromagnet into the heavy metal, the thermal spin current density polarized along the *x*-axis in the heavy metal is given by[23]

$$J_s = \Lambda \Delta T \sin\varphi \sum_k \hbar\omega_{k,\alpha} \frac{\partial f_{BE}(\omega_{k,\alpha})}{\partial T} + \Delta^2 \hbar\omega_{k,\beta} \frac{\partial f_{BE}(\omega_{k,\beta})}{\partial T}, \quad (6)$$

where $\Lambda$ is a constant depending on the interface and the density of states for the electrons in the heavy metal, $\Delta T$ is the temperature difference across the interface, $k$ is the wave vector of the one-dimensional chain, and $\Delta$ parametrizes the degree of compensation at the interface; $\Delta = 0$ corresponds to a compensated interface and $\Delta = \pm 1$ corresponds to a fully uncompensated interface where only one of the two sublattices couple to the heavy metal. The $\omega_\beta$ mode only contributes to the spin current for an uncompensated interface, reflecting the linearly polarized nature of the mode (see Supplemental Material S5 for the calculation of the spin current as a function of applied field).

The effect of the in-plane magnetic field on the pumped spin current in the heavy metal is twofold: first, the magnetic field increases the canting angle $\varphi$, causing a linear increase of the factor $\sin\varphi$ in Eq. (6)Physically, this can be interpreted by noting that each of the two sublattices pump a spin current that on average is polarized along the sublattice equilibrium direction, thus, the measured spin current is given as the projection on the *x*-axis, which is proportional to $\sin\varphi$. Second, the magnetic field changes the magnon frequencies of both magnon modes. Above the spin-flip critical field, the energy of the $\omega_\alpha$ mode and the $\omega_\beta$ mode increases with the in-plane field. This causes a decrease in the terms inside the sum in the above equation. Importantly, the increase due to the change in canting angle is proportional to $\sin\varphi \sim H$ below the critical field and constant above the critical field since the canting angle has reached its maximum at this point. In total, these two effects explain the observed peaks and saturation in SSE of $CrPS_4$/Pt at the spin-flip field.

The gap closure of the $\omega_\beta$ mode frequencies at the critical field could further increase the peak observed in the spin Seebeck effect at the critical field for systems with an uncompensated interface. However, to probe the low-frequency excitations, the temperature needs to be smaller than or comparable to the gap energy, which for $CrPS_4$ is 0.4 K in units of temperature. Therefore, a sharper peak is expected for temperatures approaching this value (see Supplemental Material S5 for details).

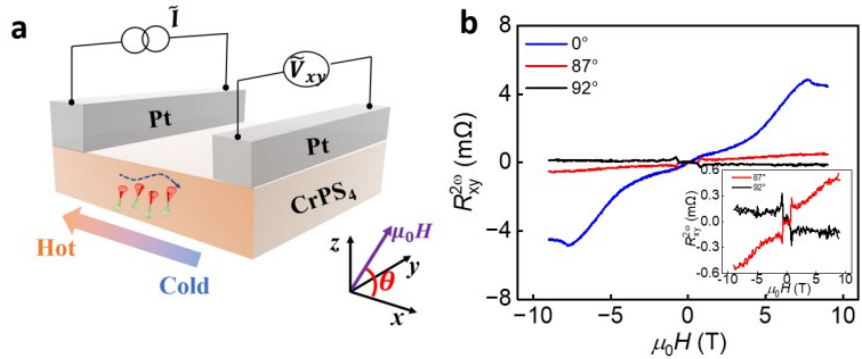

**Fig. 4 Nonlocal SSE measurement. a** Schematics of nonlocal SSE measurement. **b** Field dependence of SSE at different angles at 5 K with the applied current of 1 mA. Inset shows the field dependence SSE when the applied field is slightly off the *c*-axis (*z*-axis).

**Nonlocal SSE in CrPS$_4$/Pt.** The nonlocal configuration is further introduced to explore the SSE in CrPS$_4$/Pt as shown in Fig. 4a (see method for details). An in-plane heat gradient is created by passing current through one of the Pt strips, resulting in a nonequilibrium distribution of magnons. At the detection part, the magnon spin current is injected into Pt, which leads to the SSE. It is worth noting, in this configuration, that the temperature gradient $\nabla T$ is oriented along the *x*-axis, while the spin current $J_s$ flows along the *z*-axis, differing from the longitudinal SSE previously discussed. Fig. 4b shows the field dependence of SSE at different angles ($\theta$) at 5 K with the applied current of 1 mA. By applying the in-plane field ($\theta = 0°$), the SSE as a function of the applied field is similar to the longitudinal configuration, and a peak of SSE is also observed at the spin-flip field.

A weak SSE response occurs when the applied field is close to the z-axis, with nominal angles of $\theta = 92°$ and $87°$. Typically, the SSE should not be present when the field is directed along the *z*-axis (*c*-axis), as the parallel alignment of spin polarization $\sigma$ and spin currents $J_s$ does not generate a SSE voltage. However, a slight deviation from the *z*-axis in the direction of the applied field results in a finite value of $J_s \times \sigma$, since the spin polarization aligns with the canted magnetization. This accounts for the observed positive and negative SSE at strong positive fields when $\theta = 87°$ and $92°$, respectively. The plateau in the SSE is observed before the spin-flop transition, as there is no *x*-component of the canted magnetization. In particular, one could also find a peak of SSE at the spin-flop field, which is attributed to the divergence of spin conductance as the magnon gap closes approaching the spin-flop transition[46]. Similar effects are also observed in the longitudinal SSE configuration (see Supplemental Material S6 for details).

**Discussion**

We report evidence of the SSE in the weak interlayer exchange coupled van der Waals antiferromagnet CrPS$_4$ in contact with the heavy metal. We showed how the SSE is substantially enhanced by tuning the magnetic field. In particular, we observe a peak of SSE which shares the same temperature dependence as the spin-flip transition of CrPS$_4$ when applying magnetic field perpendicular to the *c*-axis. By considering the thermal spin current density into the heavy metal, we conclude that the SSE peak is related to the magnon mode edges as a function of the applied field across the spin-flip field.

Field-induced peaks in SSE were also observed in $Y_3Fe_5O_{12}$/Pt[47], $Lu_2BiFe_4GaO_{12}$/Pt[48], $Fe_3O_4$/Pt[49] and $Cr_2O_3$/Pt[50] bilayers. These peaks in SSE arise when the magnetic field adjusts the magnon energy to the point of anticrossing between the magnon and phonon dispersion curves, creating magnon-polarons[47]. The combined magnetoelastic excitation couples the long-lasting acoustic phonons in single crystals with the short-lived magnons, increasing the magnon lifetime and the associated SSE[48]. The SSE peak in $CrPS_4$/Pt (Ta) exhibits similar field-like behaviors, but it arises from a mechanism involving the magnon mode and spin conductance. Given that the SSE peak in $CrPS_4$/Pt (Ta) is observed at low temperatures where the phonon population is frozen, we do not expect the magnon-polarons to dominate the signal our samples.

The SSE is a sensitive tool for investigating the interfacial spin conductance and magnon population across various materials. Our findings indicate that the magnon spin transport in $CrSP_4$/Pt(Ta) can be effectively modulated through adjustments in temperature and applied magnetic field, particularly at the spin-flip field. This approach paves the way for innovative magnonic devices that utilize weakly exchange-coupled van der Waals antiferromagnetic materials.

**Method**

**Sample Preparation and Characterization**: The chemical vapor transport technique produced single crystal flakes of $CrPS_4$. Chromium (Aladdin,99.99%), red phosphorus (Aladdin,99.999%), and sulfur (Aladdin,99.999%) powders were measured in a stoichiometric ratio of 1:1:4 and combined with 5% more sulfur as transport agents. The mixed powders were sealed in a quartz tube and placed in a two-zone furnace, where the temperatures at the source and sink ends were maintained at 923 K and 823 K for a duration of 7 days. The atomic structure was analyzed using X-ray diffraction (XRD) with Cu Kα radiation (λ = 1.54056 Å). The magnetic properties were measured using a Superconducting Quantum Interference Device (SQUID). The $CrPS_4$ flakes were mechanically exfoliated from the single crystals using adhesive tape and transferred onto a SiO$_2$/Si substrate. CrPS$_4$/Pt(Ta) samples were prepared with the magnetron sputtering in a vacuum of approximately 6×10$^{-8}$ torr. The thickness of the Pt layer is 5 nm, while the Ta layer is 15 nm; 5 nm of Ta will oxidize in air, leaving 10 nm of Ta to facilitate the inverse spin Hall effect for detecting spin current generation. The Hall bar with 10 μm in width and 25 μm in length was fabricated using photolithography followed by ion beam etching. The width of the heater and the detection Pt strips are designed to be 1.4 μm and 2.3 μm, the distance of the two stipes are 1.6 μm in the nonlocal device. An atomic force microscopy image of the samples is provided in supplementary S2, showing the thickness of the CrPS$_4$ flakes to be 74 nm.

**Transport measurement**. The SSE is measured at different temperatures by varying the magnetic field in the Physical Properties Measurement System (PPMS-9T). An alternating current ranging from 0.4 to 1 mA at a frequency of 13 Hz was supplied to the Hall bar or nonlocal device using a Keithley 6221 instrument, while the transverse voltage was measured with a lock-in amplifier (SR830).

**Data availability**

The data in the main figures are provided with this paper. Other data that support the findings of this study are available from the corresponding authors upon reasonable request.

**Acknowledgements**

This work is supported by the National Key R&D Program of China (grant no. 2022YFA1203902, 2022YFA1200093), the National Natural Science Foundation of China (NSFC) (grant nos. 12241401, 12374108 and 12104052, 52373226, 52027801, 92263203), and the China-Germany Collaboration Project (M-0199), the Guangdong Provincial Quantum Science Strategic Initiative, the Fundamental Research Funds for the Central Universities, and the State Key Lab of Luminescent Materials and Devices, South China University of Technology. We acknowledge support by the German Research Foundation (CRC TRR 288 - 422213477 Project A12 and CRC TRR 173 - 268565370

**Author contributions**

S.D. and R.W. conceived the experiments. X.H. fabricated the devices. X.H., S.D., J.W. and R.W. carried out the transport and magnetic measurements. Z.C.L., Z.Y.L. and J.Y. made the single crystal samples and carried out basic characterizations. H.G.G.and A.B. contributed to the theoretical calculation. X.H., S.D., R.W., and M.K. contributed to data analysis. S.D. draft the manuscript and all authors contributed to the reviewing and revising of the manuscript. Y.H. and R.W. supervised the research and contributed to the acquisition of the financial support for the project leading to this work.

**Reference**

1. Zoui, M. A., Bentouba, S., Stocholm, J. G. & Bourouis, M. A Review on Thermoelectric Generators: Progress and Applications. *Energies* **13**, (2020).
2. Hirohata, A. et al. Review on spintronics: Principles and device applications. *J. Magn. Magn. Mater.* **509**, 166711 (2020).
3. Adachi, H., Uchida, K., Saitoh, E. & Maekawa, S. Theory of the spin Seebeck effect. *Reports Prog. Phys.* **76**, 36501 (2013).
4. Kikkawa, T. & Saitoh, E. Spin Seebeck Effect: Sensitive Probe for Elementary Excitation, Spin Correlation, Transport, Magnetic Order, and Domains in Solids. *Annu. Rev. Condens. Matter Phys.* **14**, 129–151 (2023).
5. Uchida, K. et al. Observation of the spin Seebeck effect. *Nature* **455**, 778–781 (2008).
6. Uchida, K. et al. Spin Seebeck insulator. *Nat. Mater.* **9**, 894–897 (2010).
7. Uchida, K. et al. Observation of longitudinal spin-Seebeck effect in magnetic insulators. *Appl. Phys. Lett.* **97**, 172505 (2010).
8. Cornelissen, L. J., Liu, J., Duine, R. A., Youssef, J. Ben & van Wees, B. J. Long-distance transport of magnon spin information in a magnetic insulator at room temperature. *Nat. Phys*. **11**, 1022–1026 (2015).

9. Mallick, K., Wagh, A. A., Ionescu, A., Barnes, C. H. W. & Anil Kumar, P. S. Role of spin mixing conductance in determining thermal spin pumping near the ferromagnetic phase transition in $EuO_{1-\Box}$ and $La_2NiMnO_6$. *Phys. Rev. B* **100**, 224403 (2019).

10. Geprägs, S. et al. Origin of the spin Seebeck effect in compensated ferrimagnets. *Nat. Commun.* **7**, 10452 (2016).

11. Wu, S. M. et al. Antiferromagnetic Spin Seebeck Effect. *Phys. Rev. Lett.* **116**, 97204 (2016).

12. Li, J. et al. Spin Seebeck Effect from Antiferromagnetic Magnons and Critical Spin Fluctuations in Epitaxial $FeF_2$ Films. *Phys. Rev. Lett.* **122**, 217204 (2019).

13. Gray, I. et al. Spin Seebeck Imaging of Spin-Torque Switching in Antiferromagnetic Pt/NiO Heterostructures. *Phys. Rev. X* **9**, 41016 (2019).

14. Li, J. et al. Spin current from sub-terahertz-generated antiferromagnetic magnons. *Nature* **578**, 70–74 (2020).

15. Ross, A. et al. Exceptional sign changes of the nonlocal spin Seebeck effect in antiferromagnetic hematite. *Phys. Rev. B* **103**, 224433 (2021).

16. Wu, S. M., Pearson, J. E. & Bhattacharya, A. Paramagnetic Spin Seebeck Effect. *Phys. Rev. Lett.* **114**, 186602 (2015).

17. Akopyan, A. et al. Spin Seebeck effect in $Cu_2OSeO_3$: Test of bulk magnon spin current theory. *Phys. Rev. B* **101**, 100407 (2020).

18. Chen, Y. et al. Triplon current generation in solids. *Nat. Commun.* **12**, 5199 (2021).

19. Jenni, K. et al. Chirality of magnetic excitations in ferromagnetic $SrRuO_3$. *Phys. Rev. B* **105**, L180408 (2022).

20. Rezende, S. M., Rodríguez-Suárez, R. L. & Azevedo, A. Theory of the spin Seebeck effect in antiferromagnets. *Phys. Rev. B* **93**, 14425 (2016).

21. Yamamoto, Y., Ichioka, M. & Adachi, H. Antiferromagnetic spin Seebeck effect across the spin-flop transition: A stochastic Ginzburg-Landau simulation. *Phys. Rev. B* **105**, 104417 (2022).

22. Keisuke, M. & Masahiro, S. Microscopic Theory of Spin Seebeck Effect in Antiferromagnets. *J. Phys. Soc. Japan* **93**, 34702 (2024).

23. Tang, P. & Bauer, G. E. W. Thermal and Coherent Spin Pumping by Noncollinear Antiferromagnets. *Phys. Rev. Lett.* **133**, 036701 (2024).

24. Pei, Q. L. et al. Spin dynamics, electronic, and thermal transport properties of two-dimensional CrPS$_4$ single crystal. *J. Appl. Phys*. **119**, 043902 (2016).

25. Xing, W. et al. Magnon Transport in Quasi-Two-Dimensional van der Waals Antiferromagnets. *Phys. Rev. X* **9**, 11026 (2019).

26. Qi, S. et al. Giant electrically tunable magnon transport anisotropy in a van der Waals antiferromagnetic insulator. *Nat. Commun.* **14**, 2526 (2023).

27. Lee, J. et al. Structural and Optical Properties of Single- and Few-Layer Magnetic Semiconductor CrPS$_4$. *ACS Nano* **11**, 10935–10944 (2017).

28. Peng, Y. et al. Magnetic Structure and Metamagnetic Transitions in the van der Waals Antiferromagnet CrPS$_4$. *Adv. Mater*. **32**, 2001200 (2020).

29. Ding, S. et al. Magnetic phase diagram of CrPS$_4$ and its exchange interaction in contact with NiFe. *J. Phys. Condens. Matter* **32**, 405804 (2020).

30. Calder, S. et al. Magnetic structure and exchange interactions in the layered semiconductor CrPS$_4$. *Phys. Rev. B* **102**, 24408 (2020).

31. Yang, J. et al. Layer-Dependent Giant Magnetoresistance in Two-Dimensional CrPS$_4$ Magnetic Tunnel Junctions. *Phys. Rev. Appl*. **16**, 24011 (2021).

32. Wu, R. et al. Magnetotransport Study of van der Waals CrPS$_4$/(Pt,Pd)Heterostructures: Spin-Flop Transition and Room-Temperature Anomalous Hall Effect. *Phys. Rev. Appl*. **17**, 64038 (2022).

33. Gu, K. et al. Exchange Bias Modulated by Antiferromagnetic Spin-Flop Transition in 2D Van der Waals Heterostructures. *Adv. Sci*. **11**, 2307034 (2024).

34. Son, J. et al. Air-Stable and Layer-Dependent Ferromagnetism in Atomically Thin van der Waals CrPS$_4$. *ACS Nano* **15**, 16904–16912 (2021).

35. Li, W. et al. Ultrastrong Magnon–Magnon Coupling and Chirality Switching in Antiferromagnet CrPS$_4$. *Adv. Funct. Mater*. **33**, 2303781 (2023).

36. Avci, C. O. et al. Interplay of spin-orbit torque and thermoelectric effects in ferromagnet/normal-metal bilayers. *Phys. Rev. B* **90**, 224427 (2014).

37. Behnia, K. & Aubin, H. Nernst effect in metals and superconductors: a review of concepts and experiments. *Reports Prog. Phys*. **79**, 46502 (2016).

38. Lebrun, R. et al. Tunable long-distance spin transport in a crystalline antiferromagnetic iron oxide. *Nature* **561**, 222–225 (2018).

39. Reitz, D., Li, J., Yuan, W., Shi, J. & Tserkovnyak, Y. Spin Seebeck effect near the antiferromagnetic spin-flop transition. *Phys. Rev. B* **102**, 020408 (2020).

40. Luo, R. et al. Spin Seebeck effect at low temperatures in the nominally paramagnetic insulating state of vanadium dioxide. *Appl. Phys. Lett*. **121**, 102404 (2022).

41. Kittle, C. Introduction to Solid State Physics, 4th ed. (Wiley, New York, 1966).

42. Lau, Y.-C. & Hayashi, M. Spin torque efficiency of Ta, W, and Pt in metallic bilayers evaluated by harmonic Hall and spin Hall magnetoresistance measurements. *Jpn. J. Appl. Phys*. **56**, 0802B5 (2017).

43. Kikkawa, T. et al. Longitudinal Spin Seebeck Effect Free from the Proximity Nernst Effect. *Phys. Rev. Lett*. **110**, 67207 (2013).

44. Rezende, S. M., Azevedo, A. & Rodríguez-Suárez, R. L. Introduction to antiferromagnetic magnons. *J. Appl. Phys*. **126**, 151101 (2019).

45. de Wal, D. K. et al. Long-distance magnon transport in the van der Waals antiferromagnet $CrPS_4$. *Phys. Rev. B* **107**, L180403 (2023).

46. Bender, S. A., Skarsvåg, H., Brataas, A. & Duine, R. A. Enhanced Spin Conductance of a Thin-Film Insulating Antiferromagnet. *Phys. Rev. Lett*. **119**, 56804 (2017).

47. Kikkawa, T. et al. Magnon Polarons in the Spin Seebeck Effect. *Phys. Rev. Lett*. **117**, 207203 (2016).

48. Ramos, R. et al. Room temperature and low-field resonant enhancement of spin Seebeck effect in partially compensated magnets. *Nat. Commun*. **10**, 5162 (2019).

49. Xing, W. et al. Facet-dependent magnon-polarons in epitaxial ferrimagnetic $Fe_3O_4$ thin films. *Phys. Rev. B* **102**, 184416 (2020).

50. Li, J. et al. Observation of Magnon Polarons in a Uniaxial Antiferromagnetic Insulator. *Phys. Rev. Lett*. **125**, 217201 (2020).